\documentclass[ ]{raa}            % referee version: for submission

\usepackage{graphicx,times}             %for PS/EPS graphics inclusion, new
\usepackage{natbib}
\usepackage{amssymb,amsmath}
\usepackage{mathtools}
\bibpunct{(}{)}{;}{a}{}{,}

\usepackage[a4paper=true,dvipdfm=true,pagebackref=true]{hyperref}
\hypersetup{colorlinks = true, linkcolor = green, anchorcolor = red, citecolor = blue, filecolor = red, pagecolor = red, urlcolor = red}

\begin{document}

   \title{On the error analyses of polarization measurements of the white-light coronagraph aboard ASO-S}
%   \subtitle{I. Place Your Subtitle Here}

   \volnopage{Vol.0 (20xx) No.0, 000--000}      %%preserved for Editor. DOn't remove!
   \setcounter{page}{1}          %%starting page, preserved for Editor. DOn't remove!

   \author{Li Feng
      \inst{1}
   \and Hui Li
      \inst{1}
   \and Bernd Inhester
      \inst{2}
    \and Bo Chen
      \inst{3}  
    \and Beili Ying
      \inst{1}  
    \and Lei Lu
      \inst{1} 
    \and Weiqun Gan
      \inst{1}  
   }

   \institute{Key Laboratory of Dark Matter and Space Astronomy, Purple Mountain Observatory,
Chinese Academy of Sciences, Nanjing, Jiangsu, China; {\it lfeng@pmo.ac.cn}\\
        \and
             Max-Planck-Institut f\"{u}r Sonnensystemforschung, G\"{o}ttingen, lower Saxony, Germany; {\it binhest@mps.mpg.de}\\
       \and
             Changchun Institute of  Optics, Fine Mechanics and Physics, Chinese Academy of Sciences 
\vs\no
   {\small Received~~2018 Sep 20; accepted~~2018 Oct 23}}

\abstract{The Advanced Space-based Solar Observatory (ASO-S) mission aims to explore two most spectacular eruptions in the Sun: solar flares and coronal mass ejections (CMEs), and their magnetism. For the studies of CMEs, the payload Lyman-alpha Solar Telescope (LST) has been proposed. It includes a traditional white-light coronagraph and a Lyman-alpha coronagraph which opens a new window to CME observations. 
Polarization measurements taken by white-light coronagraphs are crucial to derive fundamental physical parameters of CMEs. To make such measurements, there are two options of Stokes polarimeter which have been used by existing white-light coronagraphs for space missions. One uses a single or triple linear polarizers, the other involves both a half-wave plate and a linear polarizer. We find that the former option subjects to less uncertainty in the derived Stokes vector propagated from detector noise. The latter option involves two plates which are prone to internal reflections and may have a reduced transmission factor. Therefore, the former option is adopted as our Stokes polarimeter scheme for LST.  Based on the parameters of the intended linear polarizer(s) colorPol provided by CODIXX and the half-wave plate 2-APW-L2-012C by Altechna, it is further shown that the imperfect maximum transmittance of the polarizer significantly increases the variance amplification of Stokes vector by at least about 50\% when compared with the ideal case. The relative errors of Stokes vector caused by the imperfection of colorPol polarizer and the uncertainty due to the polarizer assembling in the telescope are estimated to be about 5\%. Among the considered parameters, we find that the dominant error comes from the uncertainty in the maximum transmittance of the polarizer.
}

   \authorrunning{Feng et al.}            %author_head in even pages
   \titlerunning{error analyses of polarization measurements}  % title_head in odd pages

   \maketitle
%% The author head (on even pages) and the title head (on odd pages) will be
%% automatically extracted from \author{} and \title{}. Whenever the title is too long,
%% you will be asked to supply a shorter one by inserting either \authorrunning{} or
%% \titlerunning{} before \maketitle. Anyway, you can specify your own heads.
%%
%%
%% Note: In the following text body of your manuscript, please note several differences from
%%       other major journals:
%% (1) \subsection{Please Capitalize the First Letter of Each Notional Word in Subsection Title}
%% (2) Please Capitalize the First Letter of Each Notional Word in all tables' captions

%
%________________________________________________ sections below
%
\section{Introduction}           %% first-level sections will be auto-capitalized

The energies of solar flares and coronal mass ejections (CMEs) are believed to originate from the solar magnetic field. The simultaneous observations of the magnetic field, flares and CMEs , and the researches on the relationship among them, are therefore of particular importance. Aiming for this major scientific objective, the Chinese solar physics community proposed the mission Advanced Space-based Solar Observatory (ASO-S)\citep{Gan:etal:2015}.
ASO-S has three payloads: the Full-disc vector MagnetoGram (FMG), the Lyman-alpha Solar Telescope (LST), and the Hard X-ray Imager (HXI) to observe vector photospheric magnetic field, CME, and flares, respectively. The mission has been in phase-B since Sep, 2017, and is scheduled to be launched in 2022 around the 25th solar activity maximum with a Sun-synchronous orbit (SSO) at an attitude of 720~km. 

LST is dedicated to the observations of the early evolution of CMEs. CME observations have often been taken by various white-light coronagraphs since the era of OSO-7 in 1970s. For white-light coronagraphs, polarization measurements are required to compute Stokes parameters, and further compute the total brightness (tB), polarized brightness (pB), etc. Based on these polarization measurements and the Thomson scattering theory for the white-light corona, physical quantities, e.g., mass, density, three dimensional locations can be derived \citep[e.g.,][]{Feng:etal:2015a, Feng:etal:2015b, Lu:etal:2017}. With LST we can not only observe the CMEs in white light but also in Lyman-alpha. Their combination allows us to derive more quantities of CMEs, e.g., their thermal properties. Nevertheless, the white-light coronagraph observations are still key ingredients in the physical diagnostics of CMEs. 

To make polarization measurements, a Stokes polarimeter needs to be included in a telescope. For  space-based white-light coronagraphs, a Stokes polarimeter can be three linear polarizers mounted on a rotating filter wheel as designed for the Large Angle Spectroscopic Coronagraph (LASCO) with three polarizers oriented at 0, 60, and -60 degrees \citep{Brueckner:etal:1995}. Instead of using three polarizers, an alternative method is to mount a linear polarizer in a hollow-core motor and rotates the polarizer to various angles as designed for the COR1 and COR2 white-light coronagraphs. With the 144-step motor design, the polarizing optic can be positioned in 2.5$^{\circ}$ increments. During normal observing operations, the polarizer mechanism will rotate 120$^{\circ}$  with an angular repeatability of better than 30 arcseconds \citep{Howard:etal:2008}. Besides linear polarizers, the Stokes polarimeter unit of white-light coronagraphs can also include other optical elements. For instance, the polarimeter assembly of the Ultraviolet Coronagraph Spectrometer (UVCS) in white light channel consists of a rotatable half-wave plate and a fixed linear polarizer \citep{Kohl:etal:1995}. 
  
In this paper, in Section 2 we first briefly introduce the LST aboard ASO-S, and its two possible options of the Stokes polarimeter for the white-light coronagraph. In Section 3, through analyzing  the effect of detector noise on Stokes vector, we evaluate the two options of the Stokes polarimeter. In Section 4, for the selected Stokes polarimeter scheme, we estimate the relative errors of Stokes vector due to the uncertainties in the polarizer parameters. The final section is conclusion and outlook. 

\section{ White-light coronagraph aboard ASO-S} 

%% Authors can give a citation as 'Michel et al. 1992'.
%% You may also use \cite, \citep and \citet for citation, and use Table~1 or Figure~1
%% and so forth. Using \ref and \label for cross-references of Tables/Figures
%% is a good way in adjusting/adding/removing text, tables or figures.
\subsection{Overview of the Lyman-alpha Solar Telescopes aboard ASO-S}

 LST consists of three instruments: Solar Disk Imager (SDI), Solar Corona Imager (SCI), and White-light Solar Telescope (WST). SDI observes the Sun up to 1.2~$\mathrm{R_{S}}$ in the Lyman-alpha line with a waveband of $121.6\pm7.5$~nm. SCI has a field of view (FOV) from 1.1 to 2.5~$\mathrm{R_{S}}$ and is a coronagraph in both the Lyman-alpha ($121.6\pm10$~nm) and white-light ($700\pm40$~nm) wavebands. WST has the same FOV as SDI but in the waveband of $360.0\pm2.0$~nm. 

The white-light and Lyman-alpha coronagraphs are equipped in the same telescope. A beam splitter is installed. The transmitted light feeds the white-light channel of SCI and goes through the Stokes polarimeter. The reflected light feeds the Lyman-alpha channel and goes through the corresponding filter. The coronal images in both the white-light and the Lyman-alpha waveband are recorded by cameras with charge coupled device (CCD) or complementary metal oxide semiconductor (CMOS) sensors.

\subsection{Mueller Matrices of the Stokes polarimeter of the white-light coronagraph}

Several calculi have been developed for analyzing polarization, including those based on the Jones matrix, coherency matrix, Mueller matrix and other matrices. Of these matrices, Mueller matrix is mostly used to characterize the polarization state change when a light beam passes a polarization element. 
A polarization state is usually defined by a Stokes vector $\mathbf{S}=(S_I, S_Q, S_U, S_V)$. For white-light coronagraphs, only linear polarization is involved. Therefore, we only consider the first three components, that is, $\mathbf{S}=(S_I, S_Q, S_U)$. From the Stokes vector, the following physical quantities can be obtained with polarimetric coronagraph measurements.
Total brightness (tB):
%\begin{linenomath*}
\begin{equation}
tB=S_I,
\label{equ:tB}
\end{equation}
%\end{linenomath*}
Polarized brightness (pB):
%\begin{linenomath*}
\begin{equation}
pB=\sqrt{S_Q^2+S_U^2},
\label{equ:pB}
\end{equation}
%\end{linenomath*}
Degree of linear polarization (DOLP):
% \begin{linenomath*}
\begin{equation}
DOLP=\frac{pB}{tB}=\frac{\sqrt{S_Q^2+S_U^2}}{S_I},
\label{equ:DOLP}
\end{equation}
%\end{linenomath*}

In the following two subsections, the Mueller matrix for the two aforementioned schemes of Stokes polarimeter are discussed. One only consists of a single rotatable linear polarizer or three fixed linear polarizers orientated at three different angles, the other consists of a rotatable half-wave plate and a fixed linear polarizer. Note that all the Mueller Matrices in this manuscript are mostly adopted or slightly modified from \citet{Bass:etal:1994}.
 
\subsubsection{Linear polarizer}
A linear polarizer is a device which produces a beam of light whose electric field vector is oscillating primarily in one plane, but still a small component in the perpendicular plane, when it is placed in an incident unpolarized light. The most basic Mueller matrix for a linear polarizer is the matrix for an ideal linear polarizer orientated at zero degree. In this case, the maximal intensity transmittance $t_{max}$ is one along the axis with zero degree, and the minimal intensity transmittance $t_{min}$ is zero along a perpendicular axis. The corresponding Mueller matrix is shown below:
%\begin{linenomath*}
\[ M_{pol}(0)=\frac{1}{2}
\left(
\begin{array}{cccc}
  1&   1& 0 & 0  \\
  1&   1& 0 & 0  \\
  0&   0& 0 & 0  \\
  0&   0& 0 & 0  
\end{array}
\right)
\]
\label{matrix:MM_ideal_pol}
%\end{linenomath*}
However, a real linear polarizer usually have $t_{max} < 1$ and $t_{min} > 0$. The Mueller matrix for the generalized case is:
%\begin{linenomath*}
\[ M_{pol}(0)=\frac{1}{2}
\left(
\begin{array}{cccc}
t_{max}+t_{min}  &  t_{max}-t_{min}   & 0                                   & 0                                  \\
t_{max}-t_{min}   &  t_{max}+t_{min}  & 0                                   & 0                                  \\
              0            &   0                        & 2\sqrt{t_{max}t_{min}}  & 0                                  \\
              0            &   0                        & 0                                   & 2\sqrt{t_{max}t_{min}}  
\end{array}
\right)
\]
\label{matrix:MM_ideal_pol1}
%\end{linenomath*}
The contrast or extinction ratio, transmittance, and diattenuation of a linear polarizer is defined as $t_{max}/t_{min}$, $t_{max}+t_{min}$, and $(t_{max}-t_{min})/(t_{max}+t_{min})$, respectively. If we write the matrix in terms of transmittance $\tau_{p}$ and diattenuation $p$,  we obtain
%\begin{linenomath*}
\[ M_{pol}(0)=\frac{\tau_{p}}{2}
\left(
\begin{array}{cccc}
              1            &   p                        & 0                                   & 0                                  \\
              p            &   1                        & 0                                   & 0                                  \\
              0            &   0                        & \sqrt{1-p^2}                   & 0                                  \\
              0            &   0                        & 0                                   & \sqrt{1-p^2}  
\end{array}
\right)
\]
\label{matrix:MM_ideal_pol2}
%\end{linenomath*} 

As we will see in the next section, for white light coronagraphs using linear polarizers, the polarizer has to be rotated to three different positions to deduce 
a Stokes vector $\mathbf{S}=(S_I, S_Q, S_U)$. It can be fulfilled either by mounting three polarizers orientated at three different positions on a filter wheel or rotating a polarizer with a hollow-core motor. When we rotate a linear polarizer or some other polarization element by an angle $\theta$, the corresponding Mueller matrix is obtained with the relationship $M(\theta)=R_M(\theta)M(0)R_M(-\theta)$, where $R_M(\theta)$ is
%\begin{linenomath*}
\[ R_M(\theta)=
\left(
\begin{array}{cccc}
              1            &   0                         & 0                                    & 0                                  \\
              0            &   \cos2\theta       & -\sin2\theta                  & 0                                  \\
              0            &   \sin2\theta        & \cos2\theta                  & 0                                  \\
              0            &   0                         & 0                                    & 1  
\end{array}
\right)
\]
\label{matrix:MM_rot}
%\end{linenomath*} 
And the linear polarizer at a position angle $\theta$ is derived accordingly with $M_{pol}(\theta)=R_M(\theta)M_{pol}(0)R_M(-\theta)$, that is, 
%\begin{linenomath*}
\[ M_{pol}(\theta)=
\left(
\begin{array}{cccc}
  t_{+}                      &   t_{-}\cos2\theta                                                                  &     t_{-}\sin2\theta                                                                & 0   \\
  t_{-}\cos2\theta   &   t_{+} \cos^22\theta+\sqrt{t_{+}^2-t_{-}^2}\sin^22\theta   &    (t_{+} -\sqrt{t_{+}^2-t_{-}^2})\cos2\theta\sin2\theta        & 0   \\
  t_{-}\sin2\theta    &   (t_{+} -\sqrt{t_{+}^2-t_{-}^2})\cos2\theta\sin2\theta         &     t_{+} \sin^22\theta+\sqrt{t_{+}^2-t_{-}^2}\cos^22\theta & 0   \\
              0                &   0                                                        & 0                                    & \sqrt{t_{+}^2-t_{-}^2}
\end{array}
\right)
\]
\label{matrix:MM_rot_pol}
%\end{linenomath*} 
where $t_{\pm}=(t_{max}{\pm}t_{min})/2$.

The polarizer that we are going to use for the white light coronagraph of LST is a colorPol polarizer from German CODIXX company. The product parameters are presented in Figure~\ref{fig:colorPol}. Our white-light coronagraph works in the waveband of $700\pm40$~nm. Read from the red and black curves, the transmittance for the selected waveband is in the range of about 0.78 to 0.82, and the contrast from $5.2\times10^5$ to $2.0\times10^4$. Such transmittance and contrast values yields the maximum transmittance $t_{max}$ from about  0.78 to 0.82, the minimum transmittance $t_{min}$ from $1.5\times10^{-6}$ to $4.0\times10^{-5}$. The derived $t_{max}$ and $t_{min}$ ranges will be further utilized in Section 3 and 4.

\subsubsection{Half-wave plate and linear polarizer}

A combination of a rotatable half-wave plate and a fixed linear polarizer is another option of the Stokes polarimeter for white light coronagraphs. A waveplate or retarder is a optical device that is used to alter the polarization state of an incident beam. A half-wave plate shifts the polarization direction of a linearly polarized light, specifically, it flips the direction around its fast axis \citep{DelToroIniesta:2003, Collett:2003}. The retardance $\delta$ of a half-wave plate is ideally to be $\delta=\pi$. For an ideal half-wave plate with its fast axis at zero degree, the corresponding Mueller matrix is 
%\begin{linenomath*}
\[ M_{HWP}(\phi=0, \delta=\pi)=
\left(
\begin{array}{cccc}
  1&   0& 0 & 0  \\
  0&   1& 0 & 0  \\
  0&   0& -1 & 0  \\
  0&   0& 0 & -1 
\end{array}
\right).
\]
\label{matrix:MM_ideal_HWP}
%\end{linenomath*}
In general cases with fast axis at an angle of $\phi$ and a retardance of $\delta$, the Mueller matrix is
%\begin{linenomath*}
\[ M_{HWP}(\phi, \delta)=
\left(
\begin{array}{cccc}
  1&   0& 0 & 0  \\
  0&   \cos^22\phi+\sin^22\phi\cos\delta & \sin2\phi\cos2\phi(1-\cos\delta)      & -\sin2\phi\sin\delta  \\
  0&   \sin2\phi\cos2\phi(1-\cos\delta)     & \sin^22\phi+\cos^22\phi\cos\delta  & \cos2\phi\sin\delta  \\
  0&   \sin2\phi\sin\delta                          &  -\cos2\phi\sin\delta                        &  \cos\delta 
\end{array}
\right).
\]
\label{matrix:MM_HWP}
%\end{linenomath*}
For the combination of a rotatable half-wave plate and a fixed linear polarizer, the resultant Mueller matrix $M(\phi, \delta, \theta)$ can be derived by $M(\phi, \delta, \theta)=M_{pol}(\theta)M_{HWP}(\phi, \delta)$. As it becomes very lengthy, we only write down the first row of $M(\phi, \delta, \theta)$. And we will see in Section 3 that to derive the Stokes vector, only the first row is involved in the calculation.
\begin{eqnarray*}
M_{11}(\phi, \delta, \theta) &=& t_{+} \\
M_{12}(\phi, \delta, \theta) &=& t_{-}\cos2\theta(\cos^22\phi+\sin^22\phi\cos\delta)+t_{-}\sin2\theta\sin2\phi\cos2\phi(1-\cos\delta) \\
M_{13}(\phi, \delta, \theta) &=& t_{-}\cos2\theta\sin2\phi\cos2\phi(1-\cos\delta)+t_{-}\sin2\theta(\sin^22\phi+\cos^22\phi\cos\delta) \\
M_{14}(\phi, \delta, \theta) &=& 0
\end{eqnarray*}

The half-wave plate that we have investigated is the 2-APW-L2-012C product from the Altechna company. Usually for a birefringent crystal, the retardance of a waveplate is denoted as $\delta=(n_e-n_o)d\,2\pi/\lambda$, and retardation as $\delta/2\pi=(n_e-n_o)d/\lambda$, where $d$ is the thickness of the waveplate, $n_e$ and $n_o$ are refractive indices of extraordinary and ordinary rays, and $\lambda$ is the wavelength. From Figure~\ref{fig:HWP}, we find that the retardation in the waveband of $700\pm40$~nm is in the range from 0.4925 to 0.51 in units of $\lambda$. Therefore, the corresponding retardance $\delta$ is from $0.985\pi$ to $1.02\pi$. For the fixed linear polarizer, the parameters $t_{max}$ and $t_{min}$ in $t_+$ and $t_-$ are the same in Section 2.2.1.

\subsection{Modulation and demodulation matrices}

Mueller matrices are used to describe the change of polarization states of incident and exiting light beams when they pass through a polarization element. However, we can not use them directly to derive the Stokes vectors that quantifying the polarization states. For intensity-measuring instruments, the $S_I$ component of a Stokes vector of the exiting light beam can be directly measured. To derive the Stokes vector (or some of its components) of the incident light beam,  we usually let the incident beam go through a polarization element with altering parameters to produce a serial measurements $S_I$ of exiting beams. As only $S_I$ of the exiting beams are involved, merely the first row of the Mueller matrix enters the calculations. One example is an incident light beam with polarization state $\mathbf{S}=(S_I, S_Q, S_U)$ goes through a linear polarizer with three altering position angles $\theta_1$, $\theta_2$, $\theta_3$.  This process can be described by
%\begin{linenomath*}
\begin{equation}
\begin{pmatrix*}[c]
S_I^{\theta_1}   \\
  S_I^{\theta_2}  \\
  S_I^{\theta_3} 
\end{pmatrix*}
=
\begin{pmatrix*}[c]
 t_{+}  &  t_{-}\cos2\theta_1         &  t_{-}\sin2\theta_1 \\
 t_{+}  &  t_{-}\cos2\theta_2         &  t_{-}\sin2\theta_2 \\
 t_{+}  &  t_{-}\cos2\theta_3         &  t_{-}\sin2\theta_3 
\end{pmatrix*}
\begin{pmatrix*}[c]
S_I\\
S_Q\\
S_U
\end{pmatrix*}
=\mathbf{O_1}\mathbf{S}.
\label{equ:ModM_O1}
\end{equation}
%\end{linenomath*}
The matrix on the right side is called the modulation matrix \citep{DelToroIniesta:Collados:2000}. By using the modulation matrix, we can set up a system of linear equations allowing a solution to the unknown $\mathbf{S}=(S_I, S_Q, S_U)$ in terms of the measured intensity $\mathbf{I}=(S_I^{\theta_1}, S_I^{\theta_2}, S_I^{\theta_3})$. If we denote the modulation matrix as $\mathbf{O}$, we have $\mathbf{I}=\mathbf{O}\mathbf{S}$. The invert matrix of $\mathbf{O}$ is called the demodulation matrix. If we designate it as $\mathbf{D}$, the unknown $\mathbf{S}$ of the incident light beam can be obtained as 
$\mathbf{S}=\mathbf{D}\mathbf{I}$. To discriminate the two Stokes polarimeter schemes, i.e., only linear polarizer(s) vs a half-wave plate plus a linear polarizer, we indicate the modulation and demodulation matrices as $\mathbf{O_1}$  and $\mathbf{D_1}$ for the former scheme, and $\mathbf{O_2}$  and $\mathbf{D_2}$ for the latter scheme.

In ideal case, $t_+=t_-=1/2$. For the three polarizers of LASCO mounted on a filter wheel, $\theta_1=-60^{\circ}$, $\theta_2=0^{\circ}$, and $\theta_3=60^{\circ}$. And in the case of the single polarizer of SECCHI/COR mounted on a hollow-core motor, $\theta_1=0^{\circ}$, $\theta_2=120^{\circ}$, and $\theta_3=240^{\circ}$. Using LASCO as an example, we have
\begin{eqnarray*}
S_I   &=& \frac{2}{3}(S_I^{-60}+S_I^{0}+S_I^{60}) \nonumber \\
S_Q &=& \frac{2}{3}(-S_I^{-60}+2S_I^{0}-S_I^{60}) \nonumber \\
S_U &=& \frac{2}{3}(-\sqrt{3}S_I^{-60}+\sqrt{3}S_I^{60}).
\end{eqnarray*}
The total brightness ($tB$) and polarized brightness ($pB$) that we often use for further analyses then can be derived using Equations~\ref{equ:tB} and \ref{equ:pB}. Consequently, 
\begin{eqnarray*}
tB &=& \frac{2}{3}(S_I^{-60}+S_I^{0}+S_I^{60}) \nonumber \\
pB&=& \frac{4}{3}\sqrt{[(S_I^{-60}+S_I^{0}+S_I^{60})^2-3(S_I^{-60}S_I^{60}+S_I^{0}S_I^{60}+S_I^{-60}S_I^{0})]}
\end{eqnarray*}
which are often seen in the references related to LASCO polarization analyses \citep[e.g.][]{Moran:Davila:2004, Lu:etal:2017}.

When using the combination of a rotatable half-wave plate and a fixed linear polarizer as the Stokes polarimeter, the modulation matrix can be similarly derived:
% \begin{linenomath*}
\begin{equation}
\begin{pmatrix*}[c]
S_I^{\phi_1}   \\
  S_I^{\phi_2}  \\
  S_I^{\phi_3} 
\end{pmatrix*}
=
\begin{pmatrix*}[c]
M_{11}(\phi_1,\delta, \theta_0)  &   M_{12}(\phi_1,\delta, \theta_0)        &  M_{13}(\phi_1,\delta, \theta_0) \\
 M_{11}(\phi_2,\delta, \theta_0)  &   M_{12}(\phi_2,\delta, \theta_0)        &  M_{13}(\phi_2,\delta, \theta_0)\\
 M_{11}(\phi_3,\delta, \theta_0)  &   M_{12}(\phi_3,\delta, \theta_0)        &  M_{13}(\phi_3,\delta, \theta_0)  
\end{pmatrix*}
\begin{pmatrix*}[c]
S_I\\
S_Q\\
S_U
\end{pmatrix*}
=\mathbf{O_2}\mathbf{S}.
\label{equ:ModM_O2full}
\end{equation}
%\end{linenomath*}
where $\phi_1$, $\phi_2$, and $\phi_3$ are the orientation angle of the half-wave plate fast axis and $\theta_0$ is the position angle of the linear polarizer. For the white-light coronagraph of LST,  $\theta_0$ is set to zero. As we mentioned before, a half-wave plate is able to mirror the polarization vector about its fast axis. For an identical incident beam,  to have equivalent effects to the linear polarizer oriented at $-60^{\circ}$, $0^{\circ}$,  and $60^{\circ}$, the fast axis of the half-wave plate is rotated to $-30^{\circ}$, $0^{\circ}$,  and $30^{\circ}$, respectively. Given such conditions, the corresponding ideal modulation scheme is 
%\begin{linenomath*}
\begin{equation}
\begin{pmatrix*}[c]
  S_I^{\phi_1}   \\
  S_I^{\phi_2}  \\
  S_I^{\phi_3} 
\end{pmatrix*}
=
\begin{pmatrix*}[c]
 0.5  &  -0.25         &  -\sqrt{3}/4  \\
 0.5  &   0.5           &  0               \\
 0.5  &   -0.25        &   \sqrt{3}/4
\end{pmatrix*}
\begin{pmatrix*}[c]
S_I\\
S_Q\\
S_U
\end{pmatrix*}.
\label{equ:ModM_O2_30}
\end{equation}
%\end{linenomath*}
The unknown $\mathbf{S}=(S_I, S_Q, S_U)$ can be derived analogously by inverting the modulation matrix. 
More generally, if we set $\theta_{0}=0$ and $\delta=\pi$ in Equation~\ref{equ:ModM_O2full}  we have
%\begin{linenomath*}
\begin{equation}
\begin{pmatrix*}[c]
S_I^{\phi_1}   \\
  S_I^{\phi_2}  \\
  S_I^{\phi_3} 
\end{pmatrix*}
=
\begin{pmatrix*}[c]
 t_{+}  &  t_{-}\cos4\phi_1         &  t_{-}\sin4\phi_1 \\
 t_{+}  &  t_{-}\cos4\phi_2         &  t_{-}\sin4\phi_2 \\
 t_{+}  &  t_{-}\cos4\phi_3         &  t_{-}\sin4\phi_3 
\end{pmatrix*}
\begin{pmatrix*}[c]
S_I\\
S_Q\\
S_U
\end{pmatrix*}
=\mathbf{O_2}\mathbf{S}.
\label{equ:ModM_O2_spec}
\end{equation}
%\end{linenomath*}
The equivalent situation with $\theta=2\phi$ in $\mathbf{O_1}$ of Equation~\ref{equ:ModM_O1} and in $\mathbf{O_2}$ of Equation~\ref{equ:ModM_O2_spec} can be derived.
Note that due to the fourth zero element in the first row of all the Mueller matrices above, we shrink the dimension of the modulation matrices from $4\times4$  to $3\times3$ and only calculate the first three components of the Stokes vector.

\section{Effect of detector noise on Stokes vector}

In this section, we compare the two aforementioned options of Stokes polarimeter for the white-light coronagraph of LST/SCI in terms of the effect of detector noise on the derived Stokes vector. For the first option, the linear polarizer(s) of SCI are positioned at $-60^{\circ}$, $0^{\circ}$, and  $60^{\circ}$, respectively. For the second option, an equivalent configuration is a rotatable half-wave plate orientated at $-30^{\circ}$, $0^{\circ}$, and  $30^{\circ}$ and a fixed linear polarizer orientated at $0^{\circ}$. In both cases, we make three measurements $S_I^p$ at three different angles $p=[-60^\circ, 0^\circ, 60^\circ]$, or $p=[-30^\circ, 0^\circ, 30^\circ]$. These three measurements have respective means $<S_I^p>$ and variances $\sigma_p^2$ due to detector noise. As shown in Section 2, the Stokes vector can be derived with $\mathbf{S}=\mathbf{D}\mathbf{I}$. The error propagation to the variances of Stokes components $\sigma_q^2  (q=I,Q,U)$ can be estimated accordingly. 
\begin{eqnarray*}
\sigma_I^2 &=& D_{11}^2\sigma_{p1}^2+D_{12}^2\sigma_{p2}^2+D_{13}^2\sigma_{p3}^2 \nonumber \\
\sigma_Q^2 &=& D_{21}^2\sigma_{p1}^2+D_{22}^2\sigma_{p2}^2+D_{23}^2\sigma_{p3}^2 \nonumber \\
\sigma_U^2 &=& D_{31}^2\sigma_{p1}^2+D_{32}^2\sigma_{p2}^2+D_{33}^2\sigma_{p3}^2 
\end{eqnarray*}
If we assume variances measured at three angles are the same, that is, $\sigma_{p1}^2=\sigma_{p2}^2=\sigma_{p3}^2=\sigma_{p}^2$, then the amplification of the detector noise induced in the Stokes components can be simplified by the sum of the squared elements in each row of the demodulation matrix $\mathbf{D}$. In cases of an ideal linear polarizer and an ideal half-wave plate in the two modulation schemes, we find that $\sigma_I^2=4/3\,\sigma_{p}^2$ and $\sigma_Q^2=\sigma_U^2=8/3\,\sigma_{p}^2$ for both $\mathbf{D}_\mathrm{1\,ideal}$ and $\mathbf{D}_\mathrm{2\,ideal}$. Therefore the amplification factor for $S_I$ is 4/3, and for $S_Q$ and $S_U$ are 8/3, respectively.

Besides using the norm of the row sum $d_q^2 = \Sigma_p(\mathbf{D}_{qp}^2)$ where $q=I,Q,U$ as a measure for the variance amplification, another measure is based on the condition number of the modulation matrices cond($\mathbf{O_1}$) and cond($\mathbf{O_2}$) \citep{Tyo:2002}. Actually the accuracy of the solution $\mathbf{S}$ to the linear equations $\mathbf{I=OS}$ depends on the condition number of the matrix $\mathbf{O}$. If $\mathbf{O}$ is well-conditioned, the computerized solution tends to be accurate. What we often use is the 2-norm for the condition number, that is, $\mathrm{cond}_2(\mathbf{O})=\lambda_{\mathrm{max}}/\lambda_{\mathrm{min}}$ where $\lambda_{\mathrm{max}}$ and $\lambda_{\mathrm{min}}$ are the maximal and minimal eigenvalues of $\mathbf{O}$. In ideal cases of $\mathbf{O}_\mathrm{1\,ideal}$ and $\mathbf{O}_\mathrm{2\,ideal}$, we find that $\mathrm{cond}_2(\mathbf{O}_\mathrm{1\,ideal})=\mathrm{cond}_2(\mathbf{O}_\mathrm{2\,ideal})=\sqrt{2}$.

\begin{figure}
   \centering
   \includegraphics[width=0.8\textwidth, angle=0]{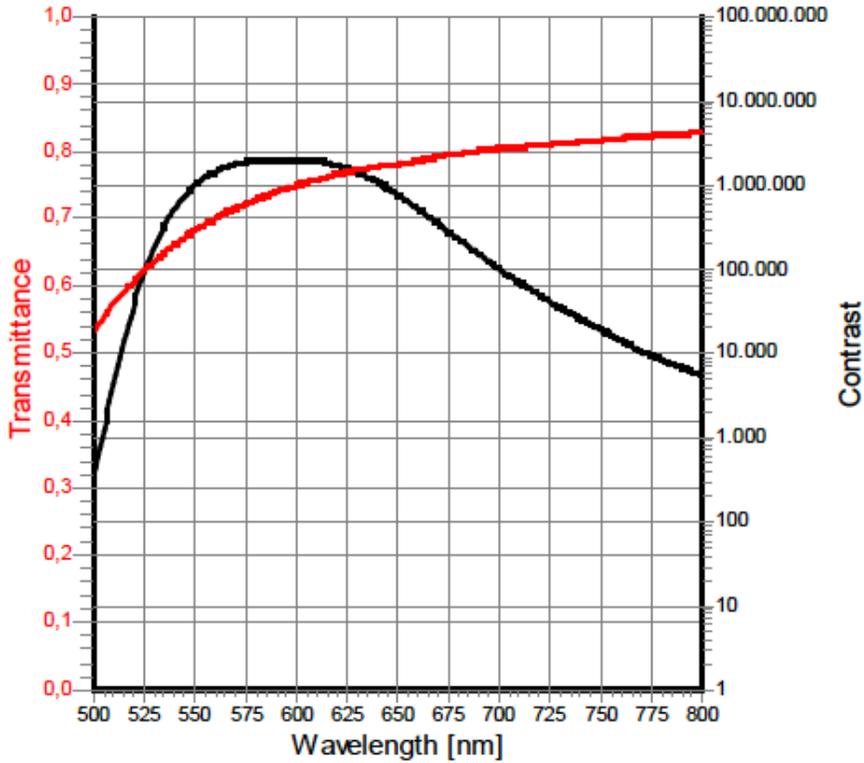}
   \caption{The transmittance and contrast as a function of wavelength for the linear polarizer product colorPol\_VIS\_BC5 are shown by red and black curves, respectively (adopted from \url{https://www.codixx.de/en/vis-visible/vis-visible-polarizer.html}).  }
   \label{fig:colorPol}
\end{figure}

\begin{figure}
   \centering
   \includegraphics[width=0.8\textwidth, angle=0]{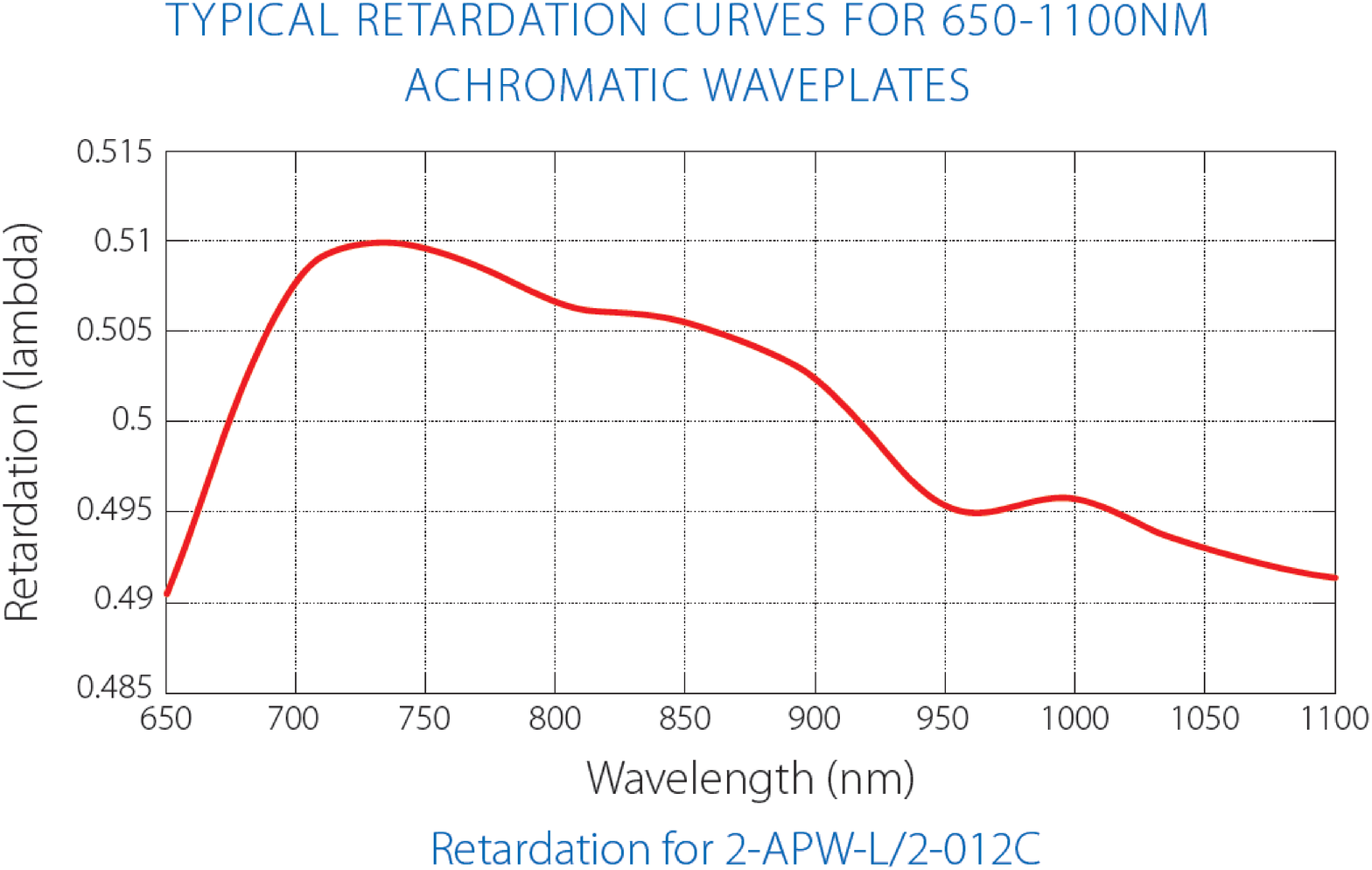}
   \caption{Retardation as a function of wavelength for the half-wave plate product 2-APW-L2-012C from the Altechna company (adopted from \url{http://www.altechna.com/product_details.php?id=878}).  }
   \label{fig:HWP}
\end{figure}

\begin{figure}
   \centering
   \includegraphics[width=\textwidth, angle=0]{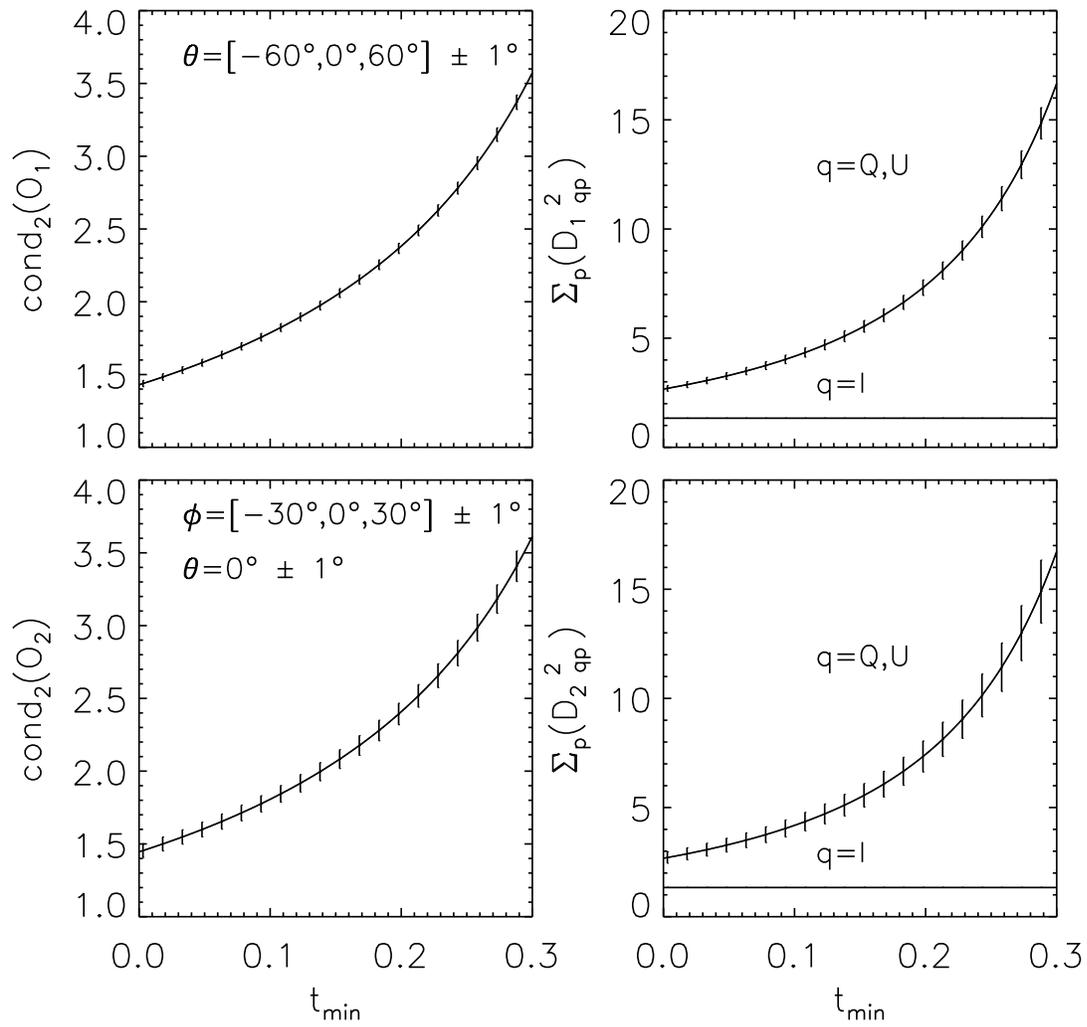}
   \caption{The condition number and variance amplification in the first (upper panels) and the second (lower panels) modulation schemes and the error bars are derived from $\pm3\sigma$ of 500 monte-carlo calculations. All the angles have a uniformly random distribution within  $\pm 1^{\circ}$ of their supposed values.}
   \label{fig:comp_pol_HWP}
\end{figure}

However, in real cases an ideal linear polarizer and an ideal half-wave plate are almost impossible to obtain. The key parameters of the products that we use for the white light coronagraph of LST/SCI are introduced in Figure~\ref{fig:colorPol} and ~\ref{fig:HWP}, respectively. Before using the specific product parameters, we have studied the variations of the condition number and amplification factors as a function of the key parameters, e.g., the minimum transmittance $t_{min}$ of a Stokes polarimeter. The upper and lower two panels of Figure~\ref{fig:comp_pol_HWP} present the condition number and variance amplification for the first and the second Stokes polarimeter schemes, respectively. The calculations are under the assumption that $t_{max}=1-t_{min}$ for the linear polarizer(s), and the uncertainty in all the orientation angles are $\pm1^\circ$. In the first case, the three angles of the linear polarizer(s) are centered at $p=[-60^\circ, 0^\circ, 60^\circ]$, and with 500 uniformly distributed numbers in the interval $[p-1^\circ, p+1^\circ]$. The resultant error bars in the upper panels of Figure~\ref{fig:comp_pol_HWP} are derived accordingly as the $3\sigma$ of these 500 calculations. We can see that the condition number and the amplification factors for $S_I$, $S_Q$ and $S_U$ start with the ideal numbers $\sqrt{2}$, 4/3, 8/3, 8/3 at $t_{min}=0$, and increase nonlinearly.  In the second case, the half-wave plate is rotated to three different positions centered at $p=[-30^\circ, 0^\circ, 30^\circ]$ in the interval $[p-1^\circ, p+1^\circ]$, and the linear polarizer is fixed at zero degree with an uncertainty of $\pm1^{\circ}$. The retardance of the half-wave plate is $\pi$ by default.  Similarly we have made 500 calculations, and the error bars in the lower panels are also $\pm3\sigma$ of these 500 calculations. The comparison of the condition number and amplification factors between the two schemes shows that they have similar mean values, but the uncertainties in the second scheme is more than two times larger than those in the first scheme. 

\begin{figure}
\centering
\vbox{
\includegraphics[trim=.1cm .5cm 0cm 1.5cm, clip, width=12.cm, height=10cm]{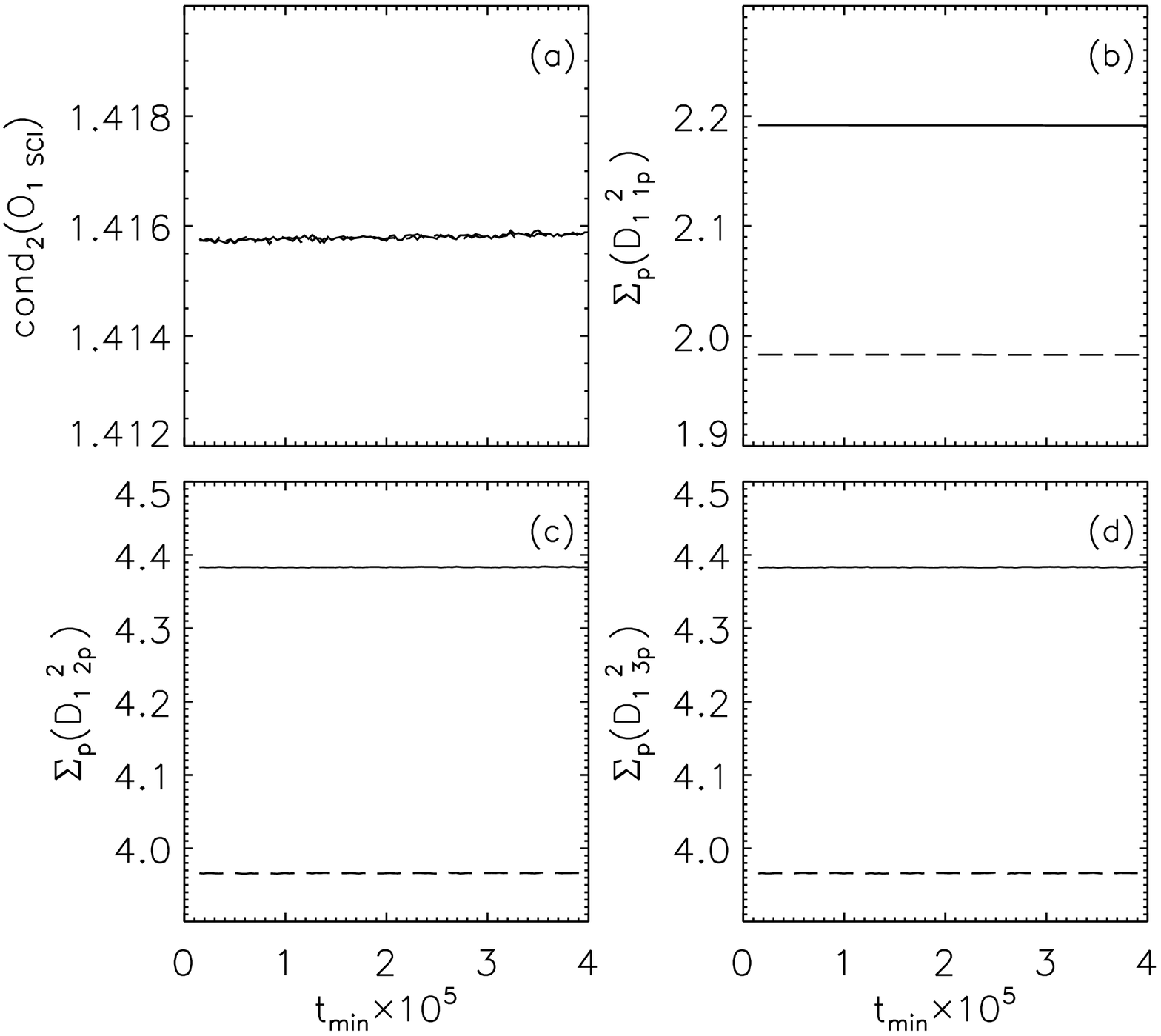}
\includegraphics[trim=.1cm .5cm 0cm 1.5cm, clip, width=12.cm, height=10cm]{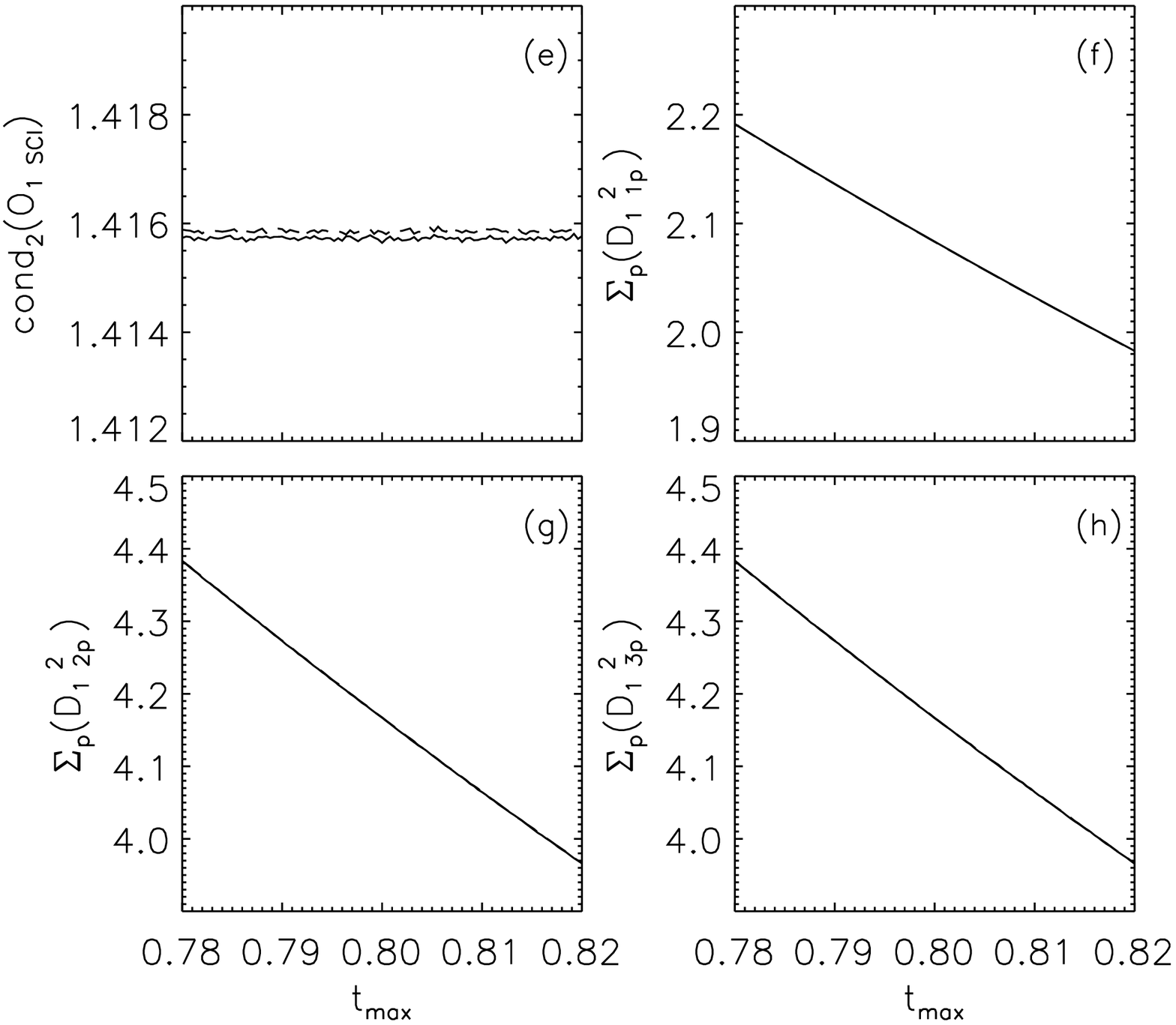}}
\caption{Condition number and variance amplifications as a function of $t_{min}$ and $t_{max}$ for using linear polarizer(s) as the Stokes polarimeter of LST/SCI. Panels (a-d) and (e-h) are the results as a function of  $t_{min}\times10^5$ and $t_{max}$. Panels (a) and (e)  present the results of the condition number. Panels (b) and (f), (c) and (g), and (d) and (h) demonstrate the variance amplification of $S_I$, $S_Q$, and $S_U$. In panels (a-d), the solid and long dashed lines represent the results for $t_{max}=0.78$ and 0.82, respectively. In panels (e-h), the solid and long dashed lines represent the results for $t_{min}=1.5\times10^{-6}$ and $4\times10^{-5}$, respectively.}
\label{fig:polar3_sci_amp}
\end{figure}

\begin{figure}
\centering
\vbox{
\includegraphics[trim=.1cm .5cm 0cm 1.5cm, clip, width=12.cm, height=10cm]{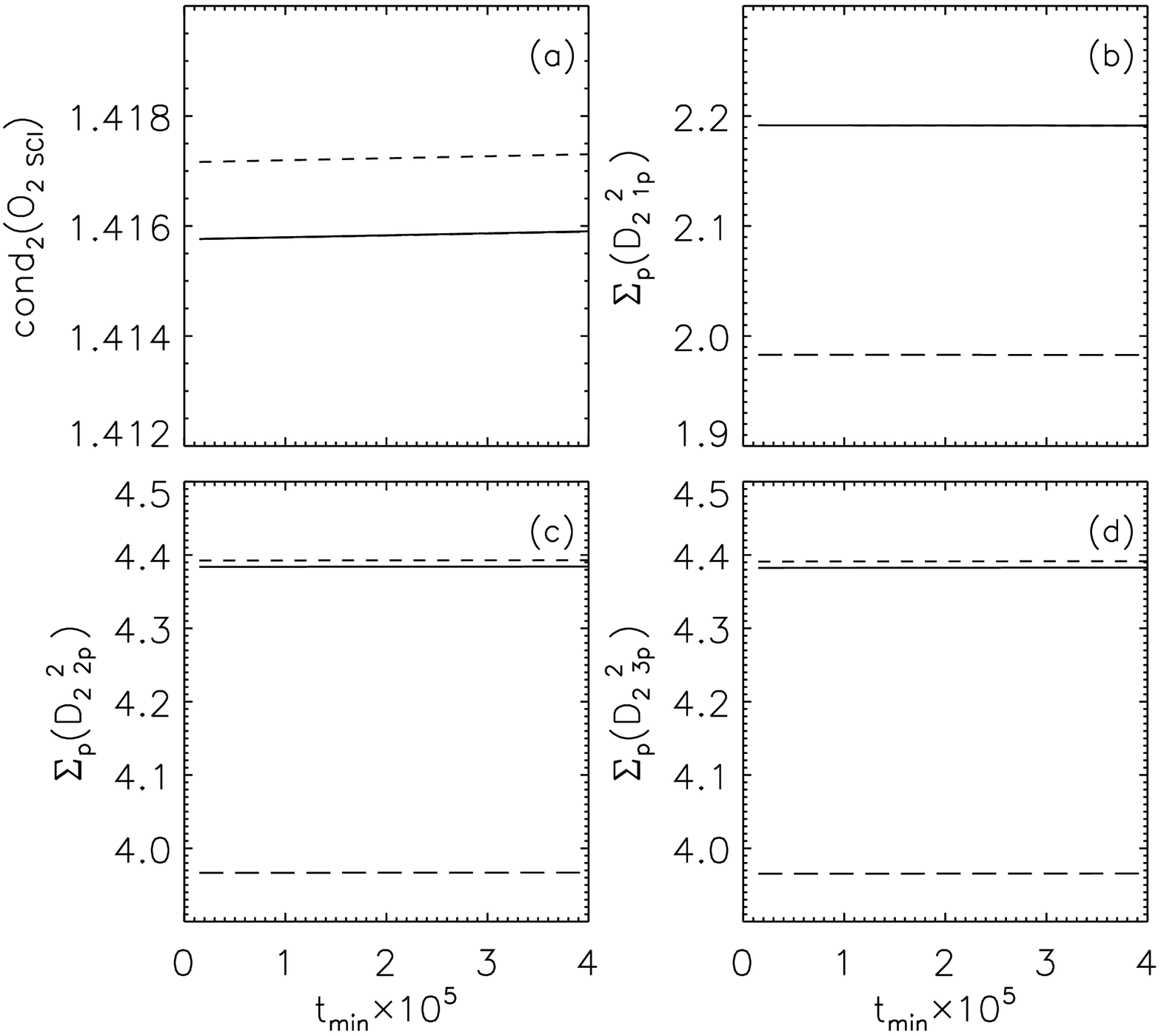}
\includegraphics[trim=.1cm .5cm 0cm 1.5cm, clip, width=12.cm, height=10cm]{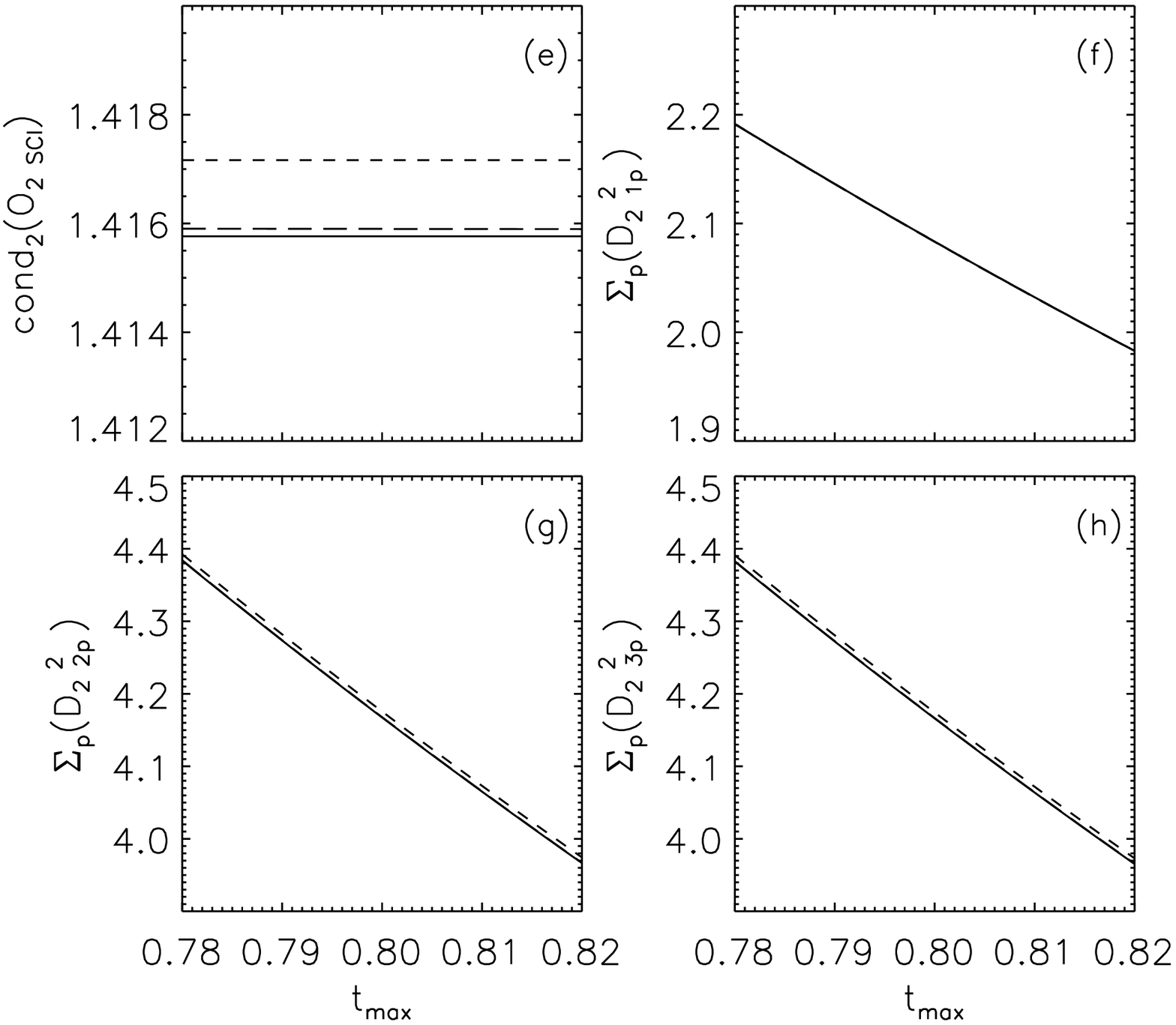}}
\caption{Condition number and variance amplifications as a function of $t_{min}$ and $t_{max}$ for using a half-wave plate and a linear polarize as the Stokes polarimeter of LST/SCI. Panels (a-d) and (e-h) illustrate the results as a function of  $t_{min}\times10^5$ and $t_{max}$. Panels (a) and (e)  present the results of the condition number. Panels (b) and (f), (c) and (g), and (d) and (h) demonstrate the variance amplification of $S_I$, $S_Q$, and $S_U$. In panels (a-d), the solid and long dashed lines delineate the results for $t_{max}=0.78$ and 0.82, $\delta=\pi$. The dashed lines represent the results for $t_{max}=0.78$ and $\delta=1.02\pi$. In panels (e-h), the solid and long dashed lines delineate the results for $t_{min}=1.5\times10^{-6}$ and $4\times10^{-5}$, $\delta=\pi$. The dashed lines represent the results for $t_{min}=4\times10^{-5}$ and $\delta=1.02\pi$. }
\label{fig:hwp_polar0_sci_amp}
\end{figure}

To evaluate the effect of the detector noise on the Stokes vector when using our selected linear polarizer and half-wave plate products described in Figure~\ref{fig:colorPol} and \ref{fig:HWP} for LST/SCI, we have computed the condition numbers and amplification factors by setting the linear polarizer $t_{min}$ in the interval $[1.5\times10^{-6}, 4.0\times10^{-5}]$, $t_{max}$ in the interval [0.78, 0.82], and half-wave plate retardance $\delta$ in the interval $[0.985\pi, 1.02\pi]$. For LST/SCI, the precision of the position angles is estimated to be up to $\pm0.1^{\circ}$. In Figure~\ref{fig:polar3_sci_amp} and \ref{fig:hwp_polar0_sci_amp}, we present the condition number and variance amplifications as a function of $t_{min}$ and $t_{max}$ for the aforementioned two Stokes polarimeter schemes. Panels (a-d) are the results as a function of  $t_{min}\times10^5$, and panels (e-h) are the results as a function of  $t_{max}$. Panels (a) and (e)  present the results of the condition number. Panels (b) and (f), (c) and (g), and (d) and (h) demonstrate the variance amplification of $S_I$, $S_Q$, and $S_U$, respectively. In the panels (a-d) of Figure~\ref{fig:polar3_sci_amp}, the solid and long dashed lines represent the results for $t_{max}=$0.78 and 0.82, respectively. In the panels (e-h) of Figure~\ref{fig:polar3_sci_amp}, the solid and long dashed lines represent the results of $t_{min}=1.5\times10^{-6}$ and $4\times10^{-5}$, respectively. Due to the very small interval of $t_{min}$, the solid and long dashed lines are overlapped onto each other. Concerning the second scheme with the half-wave plate and the linear polarizer, the designations of the lines and panels  in Figure~\ref{fig:hwp_polar0_sci_amp} are almost the same as those in Figure~\ref{fig:polar3_sci_amp}. The only difference is that for solid and long dashed lines, additionally the retardance of the half-wave plate $\delta$ is set to be $\pi$, and for dashed lines, $\delta=1.02\pi$,  $t_{max}=0.78$ or $t_{min}=4\times10^{-5}$. Because the $\pm0.1^{\circ}$ precision of the polarizer and the half-wave plate orientation angles is very high, the consequent error bars are nearly invisible, and are therefore not included in Figures~\ref{fig:polar3_sci_amp} and \ref{fig:hwp_polar0_sci_amp}. 

The linear polarizer to be used by SCI has very small $t_{min}$. Thus by comparing Figures~\ref{fig:polar3_sci_amp} and \ref{fig:hwp_polar0_sci_amp} and the ideal values in Figure~\ref{fig:comp_pol_HWP} , we can evaluate the effects of $t_{max}$ and $\delta$ on the condition numbers and variance amplifications. The lower $t_{max}$ decreasing from unity to [0.78, 0.82] lifts the condition number from $\sqrt{2}$ to about 1.416, and the imperfection of the half-wave plate with $\delta=1.02\pi$ further increases the condition number to about 1.417. The elevation of variance amplifications is more prominent.
The ideal numbers for  $S_I$, $S_Q$, and $S_U$ are 4/3, 8/3 and 8/3. After considering the imperfect $t_{max}$ and $\delta$, these numbers increased significantly.  For instance, the amplification factor for  $S_I$ increases from 4/3 to the interval [1.98, 2.19], and  for $S_U$ and $S_Q$ elevates from 8/3 to the interval [3.97, 4.39]. The increment of the amplification factors ranges from 49\% to 65\%.

Actually according to Equation~\ref{equ:ModM_O1} and ~ \ref{equ:ModM_O2_spec}, the errors introduced by uncertainties of the assembly parameters in these two schemes differs. $\phi$ has a factor 4 so it must be known twice as good as the polarizer angle $\theta$. In addition, the HWP assembly has one more parameter. The retardation $\delta$ in Equation~\ref{equ:ModM_O2full} could be an additional source of uncertainty. Moreover, there is an additional disadvantage of the HWP assembly. It needs two plates which are prone to internal reflections and may have a reduced transmission factor. Therefore, for LST/SCI we tend to adopt the first scheme with only linear polarizers involved.

\section{Errors of Stokes vector due to the uncertainties in the polarizer parameters} 

This section is dedicated to the calculation of errors of Stokes vector due to the uncertainties in the polarizer parameters: $t_{max}$, $t_{min}$, and $\theta$ for LST/SCI. Because $\Delta t_\mathrm{max}=0.04$, $\Delta t_\mathrm{min}=4\times10^{-5}$, and $\Delta \theta=1^{\circ}=0.0017\mathrm{rad}$ are small, we can linearize \citep{Tyo:2002}
\begin{equation}
\frac{d\mathbf S} {dt_\mathrm{max}}  =  \frac{d}{dt_\mathrm{max}} \mathbf{D_1}
\begin{pmatrix*}[c]
S_I^{-60}  \\
S_I^{0}  \\
S_I^{60} 
\end{pmatrix*}
,\quad
\frac{d\mathbf S} {dt_\mathrm{min}}  =  \frac{d}{dt_\mathrm{min}} \mathbf{D_1}
\begin{pmatrix*}[c]
S_I^{-60}  \\
S_I^{0}  \\
S_I^{60} 
\end{pmatrix*}
,\quad
\frac{d\mathbf S} {d\theta}  =  \frac{d}{d\theta} \mathbf{D_1}
\begin{pmatrix*}[c]
S_I^{-60}  \\
S_I^{0}  \\
S_I^{60} 
\end{pmatrix*}
.
\label{equ:s_deriv}
\end{equation}
Then the variance in $S_q (q=I,Q,U)$ due to uncertainties $\Delta t_\mathrm{max}$, $\Delta t_\mathrm{min}$, and $\Delta \theta$ can be calculated by 
\begin{equation}
\sigma^2_q=(\frac{dS_q}{dt_\mathrm{max}}\Delta t_\mathrm{max})^2+(\frac{dS_q}{dt_\mathrm{min}}\Delta t_\mathrm{min})^2+(\frac{dS_q}{d\theta}\Delta\theta)^2.
\label{equ:error_sigq}
\end{equation}
Because we can not obtain the derivatives of $\mathbf{D_1}$ analytically,  we reformat it as the derivative of $\mathbf{O_1}$. For instance, the derivative with respect to $\theta$ can be rewritten as:

\begin{equation}
\frac{d\mathbf{S}}{d\theta}=\frac{d}{d\theta} \mathbf{D_1}
\begin{pmatrix*}[c]
S_I^{-60}  \\
S_I^{0}  \\
S_I^{60} 
\end{pmatrix*}
=\frac{d}{d\theta}(\mathbf{D_1})\mathbf{O_1S}
=-\mathbf{D_1}\frac{d}{d\theta}(\mathbf{O_1})\mathbf{S}
\end{equation}
Deriving the derivatives of $\mathbf{O_1}$ is straight-forward. The results are
\begin{equation}
\frac{d}{dt_\mathrm{max}}
\begin{pmatrix*}[c]
 t_{+}  &  t_{-}\cos2\theta_1         &  t_{-}\sin2\theta_1 \\
 t_{+}  &  t_{-}\cos2\theta_2         &  t_{-}\sin2\theta_2 \\
 t_{+}  &  t_{-}\cos2\theta_3         &  t_{-}\sin2\theta_3 
\end{pmatrix*}
=
\frac{1}{2}
\begin{pmatrix*}[c]
1 & \cos2\theta_1 & \sin2\theta_1 \\
1 & \cos2\theta_2 & \sin2\theta_2 \\
1 & \cos2\theta_3 & \sin2\theta_3 
\end{pmatrix*}
\end{equation}
\begin{equation}
\frac{d}{dt_\mathrm{min}}
\begin{pmatrix*}[c]
 t_{+}  &  t_{-}\cos2\theta_1         &  t_{-}\sin2\theta_1 \\
 t_{+}  &  t_{-}\cos2\theta_2         &  t_{-}\sin2\theta_2 \\
 t_{+}  &  t_{-}\cos2\theta_3         &  t_{-}\sin2\theta_3 
\end{pmatrix*}
=
\frac{1}{2}
\begin{pmatrix*}[c]
1 & -\cos2\theta_1 & -\sin2\theta_1 \\
1 & -\cos2\theta_2 & -\sin2\theta_2 \\
1 & -\cos2\theta_3 & -\sin2\theta_3 
\end{pmatrix*}
\end{equation}

\begin{equation}
\frac{d}{d\theta}
\begin{pmatrix*}[c]
 t_{+}  &  t_{-}\cos2\theta_1         &  t_{-}\sin2\theta_1 \\
 t_{+}  &  t_{-}\cos2\theta_2         &  t_{-}\sin2\theta_2 \\
 t_{+}  &  t_{-}\cos2\theta_3         &  t_{-}\sin2\theta_3 
\end{pmatrix*}
=
\begin{pmatrix*}[c]
0 & -2t_{-}\cos2\theta_1 & 2t_{-}\sin2\theta_1 \\
0 & -2t_{-}\cos2\theta_2 & 2t_{-}\sin2\theta_2 \\
0 & -2t_{-}\cos2\theta_3 & 2t_{-}\sin2\theta_3 
\end{pmatrix*}
\end{equation}

\begin{figure}
  \centering
   \includegraphics[width=8cm, height=20cm, angle=0]{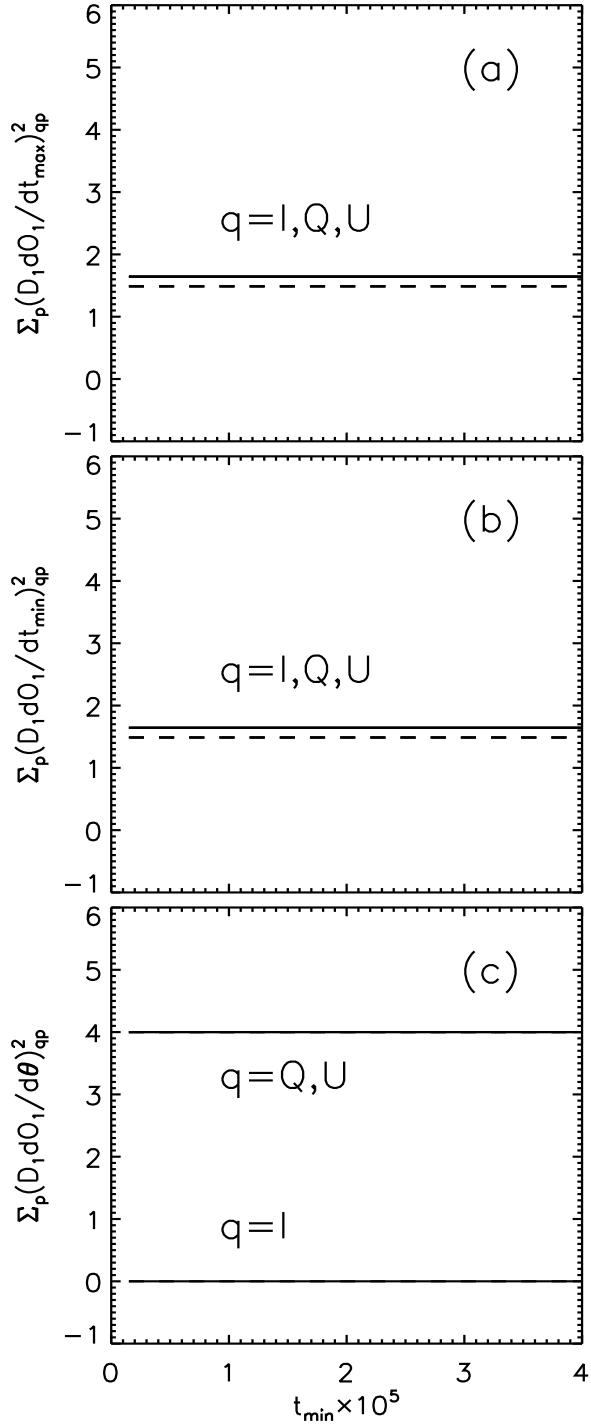}
   \caption{The sum of the squared elements in each row of $\mathbf{D_1}\frac{d}{dt_\mathrm{max}}(\mathbf{O_1})$, $\mathbf{D_1}\frac{d}{dt_\mathrm{min}}(\mathbf{O_1})$, $\mathbf{D_1}\frac{d}{d\theta}(\mathbf{O_1})$ as a function of $t_{min}\times10^5$. The solid and dashed lines correspond to the results for $t_{max}=0.78$ and $t_{max}=0.82$, respectively.}
   \label{fig:error_IQU}
\end{figure}

In Figure~\ref{fig:error_IQU} we plot the sum of the squared elements in any row of $\mathbf{D_1}\frac{d}{dt_\mathrm{max}}(\mathbf{O_1})$, $\mathbf{D_1}\frac{d}{dt_\mathrm{min}}(\mathbf{O_1})$, $\mathbf{D_1}\frac{d}{d\theta}(\mathbf{O_1})$ for $t_{min}$ in the interval $[1.5\times10^{-6}, 4.0\times10^{-5}]$, $t_{max}$ in the interval [0.78, 0.82], $\theta$ in the interval $[-60,0,60]\pm0.1^{\circ}$. In panels (a) and (b), the solid lines and dashed lines represent the $\Sigma_p(\mathbf{D_1}d\mathbf{O_1}/dt_\mathrm{max})_{qp}^2$, $\Sigma_p(\mathbf{D_1}d\mathbf{O_1}/dt_\mathrm{min})_{qp}^2$ as a function of $t_{min}\times10^5$ for $t_{max}=0.78$ and $t_{max}=0.82$, respectively. In the very small range of $t_{min}$, we find that both sums for different $q=I,Q,U$ are identical and have almost no variation, and there is only 12\% difference between the results for $t_{max}=0.78$ and $t_{max}=0.82$. In panel (c), the uncertainty $\Sigma_p(\mathbf{D_1}d{\mathbf{O_1}/d\theta})_{qp}^2$ in $\theta$ depends neither on $t_{min}$ nor $t_{max}$, as the solid lines and dashed lines are superimposed onto each other and there is no variation along $t_{min}$.  However, the uncertainty for $q=Q,U$ and for $q=I$ are quite different. Note that when we increase $\Delta \theta$ from $0.1^{\circ}$ to $10^{\circ}$ or even some larger numbers, Figure~\ref{fig:error_IQU} almost has no change. Therefore, a deviation of the polarizer orientations from the correct angles does not seem to enhance 
$\Sigma_p(\mathbf{D_1}d\,\mathbf{O_1}/dt_\mathrm{max})_{qp}^2$, $\Sigma_p(\mathbf{D_1}d\,\mathbf{O_1}/dt_\mathrm{min})_{qp}^2$, and $\Sigma_p(\mathbf{D_1}d\,\mathbf{O_1}/d\theta)_{qp}^2$.

An estimate of the induced maximum relative uncertainty in the Stokes parameters is obtained by multiplying the polarizer parameter errors to the respective ordinate value in In Figure~\ref{fig:error_IQU}. For given $\Delta t_\mathrm{max}=0.04$, $\Delta t_\mathrm{min}=4\times10^{-5}$, and $\Delta \theta=0.0017\mathrm{rad}$, according to Equation~\ref{equ:error_sigq}, we find that the relative errors $\sigma_I/S_I \approx \sigma_Q/S_Q \approx \sigma_U/S_U \approx 5\%$, and the major error comes from $\Delta t_\mathrm{max}$. 

\begin{figure}
  \centering
   \includegraphics[width=16cm, height=11cm, angle=0]{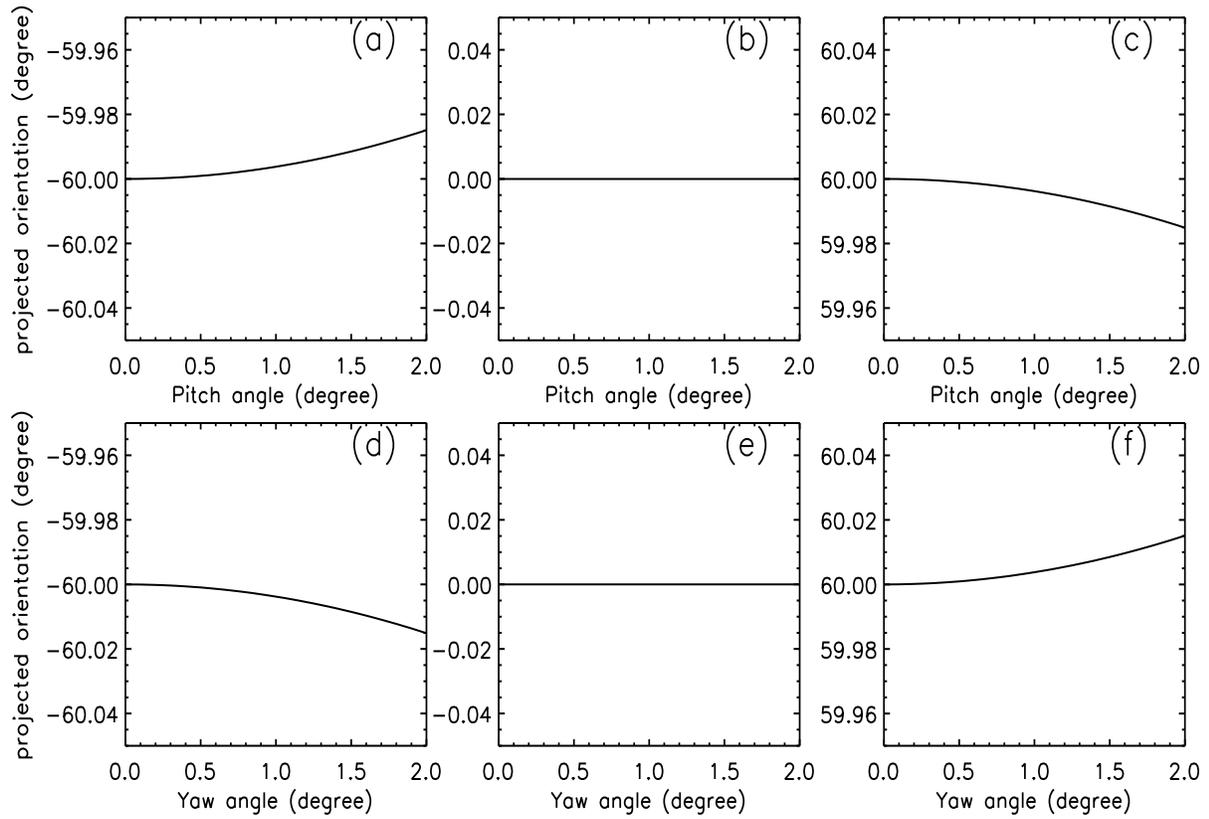}
   \caption{Projected orientation angles from $-60^\circ, 0^\circ, 60^\circ$ measured in the plane defined by the pitch and yaw angles onto the plane with zero pitch and yaw angles. }
   \label{fig:orien_pitch_yaw}
\end{figure}

The last but not least error source to consider is the pitch and yaw angles of the linear polarizer mounted on a filter wheel or a hollow-core motor. We assume the maximal pitch and yaw angles are $2^\circ$ and project the polarizer orientation angle measured in the plane defined by the pitch and yaw angles onto the plane with zero pitch and yaw angles. Note that the designed three orientation angles are intended to be $-60^\circ, 0^\circ, 60^\circ$ in the plane defined by the pitch and yaw angles. The projected orientation angles as a function of pitch and yaw angles are displayed in Figure~\ref{fig:orien_pitch_yaw}. The upper and lower panels show the results as a function of pitch angle and yaw angle, respectively. We find that in the projected plane with zero pitch and yaw angles, the maximal deviations from $-60^\circ, 0^\circ, 60^\circ$ are $0.015^\circ, 0^\circ, 0.015^\circ$, respectively. Therefore, the errors induced by the pitch and yaw angles with a maximum of $2^\circ$ can be negligible.

\section{Conclusions and Outlook}
For studies of CMEs in the corona above the solar limb, ASO-S carries a white-light and Lyman-alpha coronagraph. This paper is dedicated to the polarization measurements taken by the white-light coronagraph. There are two options of Stokes polarimeter which are often implemented for white-light coronagraphs. One consists of either a single rotatable linear polarizer mounted on a hollow-core motor or three fixed linear polarizers with different orientation angles mounted on a filter wheel. The other consists of a  rotatable half-wave plate and a fixed linear polarizer. For these two schemes of Stokes polarimeter, we have calculated their corresponding Mueller, modulation, and demodulation matrices for further analyses.

We have compared the effect of detector noise on the Stokes vector in terms of the condition number of modulation matrix and the amplification factor of the measurement variance to the variance of Stokes vector. It shows that both options of Stokes polarimeter have similar mean condition number and amplification factors, but the scheme using only linear polarizer(s) is subject to less uncertainty caused by the imperfection of orientation angles. 
Moreover, the latter option involves two plates which are prone to internal reflections and may have a reduced transmission factor. Therefore, we intend to adopt the first scheme for the white-light coronagraph of LST/SCI. Within the first scheme, we find after experimenting in the lab that using three linear polarizers orientated at three different angles installed in a filter wheel is more reliable than using a single rotatable linear polarizer mounted in a hollow-core motor.

We further calculate the effect of detector noise on the Stokes vector using the parameters of the linear polarizer product from CODIXXX and the half-wave plate product from Altechna. The minimum transmittance $t_{min}$ is close to the ideal situation, whereas the maximum transmittance $t_{max}$ is significantly reduced. Such a decrease slightly increases the condition number and greatly elevates the amplification factors by at least 50\% from their ideal levels. The additional imperfection of the retardance of the half-wave plate has only a slightly further increment to the amplification factors. 

Finally we estimate the relative errors of Stokes vector due to uncertainties in the polarizer parameters for the first scheme of the Stokes polarimeter which will be used by LST/SCI. It is found that the relative errors are about 5\% and the major error comes from the large uncertainty in $t_{max}$. $t_{min}$ and $\theta$ have very limited contributions. The error induced by the assembling pitch and yaw angles of the linear polarizer(s) within $2^\circ$ is evaluated to be negligible. The current calculations are based on the product parameters provided by the company. In the future, these parameters will be measured in the laboratory which may narrow down the error estimate of our polarization measurements.

\begin{acknowledgements}

We thank Zhongquan Qu and Junfeng Hou for useful discussions. This work is supported by NSFC
grants (11522328, 11473070, 11427803, U1731241), by CAS Strategic Pioneer Program on Space Science, Grant No. XDA15010600, XDA15052200, XDA15320103, and XDA15320301, and National key research and development program 2018YFA0404202.

\end{acknowledgements}

\bibliographystyle{raa}
%\bibliography{asos}
%\begin{thebibliography}{99}
%% you can type \apj for ApJ, \aap for A&A, \apss for Ap&SS, etc. Please consult
%% the macro chjaa.cls. You can also find them in aasguide.tex (AASTeX for ApJ, AJ, PASP)
%% Please follow the format of ChJAA's reference list
%\end{thebibliography}

\label{lastpage}

\end{document}